\documentclass[letterpaper,10pt]{article} 

\usepackage{opticameet3} 

\newcommand\authormark[1]{\textsuperscript{#1}}

\usepackage{amsmath,amssymb}
\usepackage{braket}

\usepackage[colorlinks=true,bookmarks=false,citecolor=blue,urlcolor=blue]{hyperref} 
\begin{document}

\title{Galois Symmetries in the Classification and Quantification of Quantum Entanglement}


\author{Bilal Benzimoun,\authormark{1,*}, Abdelali Sajia \authormark{2}}
\address{\authormark{1} Department of Physics, Clark University, Worcester, Massachusetts 01610, USA \\
\authormark{2} Center for Quantum Science and Engineering, and Department of Physics, Stevens Institute of Technology, Hoboken, New Jersey 07030, USA }
\email{\authormark{*}bbenzimoun@clarku.edu}

\begin{abstract}
Quantum entanglement, a cornerstone of quantum mechanics, remains challenging to classify, particularly in multipartite systems. Here, we present a new interpretation of entanglement classification by revealing a profound connection to Galois groups, the algebraic structures governing polynomial symmetries. This approach not only uncovers hidden geometric relationships between entangled quantum states and polynomial roots but also introduces a method for quantifying entanglement in multi-qubit symmetric states. By reframing the classification of GHZ, $W$, and separable states within the structure of Galois symmetries, we establish a previously unrecognized hierarchy in their entanglement properties. This work bridges the mathematical elegance of Galois theory with the complexities of quantum mechanics, opening pathways for advances in quantum computing and information theory.\\
\end{abstract}

\section{Introduction}
Quantum entanglement lies at the heart of the quantum world, enabling phenomena such as superposition and non-locality, which defy classical intuition \cite{Horodecki2009}. As the foundation of quantum technologies—including quantum computing, cryptography, and communication—entanglement is both a fundamental and practical resource \cite{Nielsen2000, Ali}. However, despite its central role, the classification and measurement of entanglement, particularly in multipartite quantum systems, remains an unsolved challenge \cite{Guhne2009}. The complexity of these quantum correlations grows exponentially with the number of qubits, making it increasingly difficult to classify or quantify entanglement in a comprehensive manner \cite{Amico2008}.

In this work, we offer a new perspective on this long-standing problem by linking quantum entanglement with the algebraic structure of Galois groups. Traditionally, Galois theory has provided a deep understanding of symmetries in polynomial equations, revealing the hidden relationships between the roots of these equations \cite{Stewart2004}. Here, we extend this powerful framework to the quantum realm, showing that the symmetries captured by Galois groups offer a natural means of classifying and quantifying entanglement in multi-qubit systems.

Specifically, we demonstrate that multipartite entangled states—including the well-known GHZ \cite{Greenberger1990} and $W$ states \cite{Dur2000}—can be reinterpreted through the lens of Galois symmetries, uncovering an inherent hierarchy in their entanglement properties. More importantly, this connection allows us to propose a method for measuring entanglement in multi-qubit symmetric states\cite{Book1}, addressing a key gap in current quantum theory \cite{Wei2003}. By revealing how algebraic structures underpin quantum correlations, our approach not only advances the theoretical understanding of entanglement but also opens up new possibilities for practical quantum information applications \cite{Lamata2007}.

This intersection of algebra and quantum mechanics represents a significant step forward, offering insights that extend beyond the classification of quantum states. It invites further exploration into how mathematical frameworks like Galois theory might inform our understanding of other fundamental quantum phenomena, with potential implications for developing more robust quantum systems and novel computational architectures \cite{Bengtsson2006}.

\section{Correspondence between spin coherent states and symmetric multi-qubit states}

As early as 1932, the Italian physicist Ettore Majorana proposed a profound connection between the quantum states of spin systems and geometric representations on the Bloch sphere~\cite{Majorana1932, Bloch1946}. Specifically, Majorana showed that any pure symmetric state of an $N$-qubit system can be represented by a set of $N$ spinors (single-qubit states), where each spinor corresponds to a point on the Bloch sphere. This geometric representation, now known as the \textit{Majorana representation}, allows for a visual and intuitive understanding of complex quantum states and their entanglement properties.

The essence of Majorana's connection lies in expressing a symmetric multi-qubit state as a symmetrized product of single-qubit states (spinors). Because symmetric states are invariant under the permutation of qubits, they can be constructed by summing over all permutations of individual spinor states. This symmetrization reflects the indistinguishability and collective behavior of the qubits in such states.

Mathematically, this connection is expressed as:

\begin{equation}
    \ket{\Psi_{sym}} = \mathcal{N} \sum_{\sigma \in S_N} \left\{ \ket{\epsilon_{\sigma_{(1)}}, \epsilon_{\sigma_{(2)}}, \ldots, \epsilon_{\sigma_{(N)}}} \right\},
\end{equation}
where
\begin{equation}
    \ket{\epsilon_l} = \cos{\left( \frac{\beta_l}{2} \right)} e^{-i \frac{\alpha_l}{2}} \ket{0} + \sin{\left( \frac{\beta_l}{2} \right)} e^{i \frac{\alpha_l}{2}} \ket{1}, \quad l = 0,1,2,\ldots,N.
\end{equation}

In equation (1), the summation is over the set of all \( N! \) permutations of the qubits \( \left\{ \ket{\epsilon_l} \right\} \), and \( \mathcal{N} \) is the normalization factor.

\noindent Every symmetric state of \( N \) qubits can be uniquely expressed in the Dicke basis, formed by the \( N + 1 \) joint eigenstates \( \left\{ \ket{\frac{N}{2}, l - \frac{N}{2}} \right\} \), with \( l = 0, 1, 2, \ldots, N \), of the collective operators \( \hat{S_z} = \sum_{i=1}^N \hat{\sigma}^z_i \) and \( \hat{S}^2 \), where \( \hat{S} = \sum_{i=1}^N \hat{\sigma}_i \) \cite{Dicke1954}. These states are given by:

\begin{equation}
    \ket{\frac{N}{2}, l - \frac{N}{2}} = \sqrt{\frac{l!(N-l)!}{N!}} \left( \ket{\underbrace{0,0,\ldots,0}_{l}, \underbrace{1,1,\ldots,1}_{N-l}} + \text{permutations} \right).
\end{equation}

\noindent The cases \( l = 0 \) and \( l = N \) correspond to the states \( \ket{\frac{N}{2}, -\frac{N}{2}} \equiv \ket{1,1,\ldots,1} \) and \( \ket{\frac{N}{2}, \frac{N}{2}} \equiv \ket{0,0,\ldots,0} \), respectively.

\noindent For \( N = 2 \), equation (1) reduces to:

\[
\ket{\Psi_{sym}} = \mathcal{N}' \left( \ket{\epsilon_1, \epsilon_2} + \ket{\epsilon_2, \epsilon_1} \right),
\]
which describes a pure symmetric state corresponding to a spin-1 particle \cite{Arecchi1972}. In the Dicke basis \( \left\{ \ket{j=1,m}, m = -1,0,1 \right\} \), each \( m \) corresponds to the state of a spin-1 particle:
\[
\left\{ \ket{j=1,m=-1} = \ket{0,0}, \ket{j=1,m=0} = \frac{1}{\sqrt{2}}(\ket{1,0} + \ket{0,1}), \ket{j=1,m=1} = \ket{1,1} \right\}.
\]

An arbitrary pure state of a spin-\( j \) can be expressed as a symmetric state in the Dicke basis as follows. In the rest of the discussion, we will replace the coefficients \( c_{j+m} \) with \( c_l \), where \( l = 0,1,\ldots,N \).

\begin{equation}
    \ket{\Psi} \equiv \sum_{m=-j}^j c_{j+m} \Ket{j,m} = \mathcal{N} \sum_{\sigma \in S_N} \left\{ \ket{\epsilon_{\sigma_{(1)}}, \epsilon_{\sigma_{(2)}}, \ldots, \epsilon_{\sigma_{(N)}}} \right\}
    = \sum_{l=0}^N c_l \ket{\frac{N}{2} = j, l - \frac{N}{2}}.
\end{equation}

In this way, the state of a spin-\( j \) particle can be decomposed into \( 2j \) spin-\( \frac{1}{2} \) particles in the Dicke basis, represented by the set \( \left\{ \ket{\frac{N}{2}, l - \frac{N}{2}} \right\} \). This set can be viewed as a symmetric state of \( N = 2j \) spin-\( \frac{1}{2} \) particles, which is geometrically represented by \( 2j \) points on the Bloch sphere, known in the literature as Majorana stars \cite{Bacry1974, Bloch1946, Ribeiro2007}.

\noindent There is a simple way to express the coefficients \( c_l \) in terms of the orientations of the Majorana spinors \( \left( \alpha_l, \beta_l \right) \) \cite{Hannay1998}. A rotation \( R \otimes R \ldots \otimes R \) on the symmetric state transforms it into another symmetric state. Choosing \( R^{-1}\left( \alpha_s, \beta_s \right) \otimes R^{-1}\left( \alpha_s, \beta_s \right) \ldots \otimes R^{-1}\left( \alpha_s, \beta_s \right) \), where \( \alpha_s, \beta_s \) correspond to the orientation of any one of the spinors, we have:

\begin{equation}
    \bra{1,1,\ldots,1} R^{-1}\left( \alpha_s, \beta_s \right) \otimes \ldots \otimes R^{-1}\left( \alpha_s, \beta_s \right) \ket{\Psi_{sym}} = 0.
\end{equation}

This is because the rotation \( R^{-1}_s \otimes \ldots \otimes R^{-1}_s \) transforms one of the spinors \( \ket{\epsilon_s} \) with the orientation angles \( \left( \alpha_s, \beta_s \right) \) to \( \ket{0} \). For \( N \) rotations \( R^{-1}_s, s=1,2,\ldots,N \), we get:

\begin{equation}
    \sum_{l=0}^N c_l \bra{\frac{N}{2}, -\frac{N}{2}} \mathcal{R}^{-1}_s \ket{\frac{N}{2}, l - \frac{N}{2}} = 0,
\end{equation}
where \( \mathcal{R}^{-1}_s \equiv R^{-1}_s \otimes \ldots \otimes R^{-1}_s \). This expression can be written as:

\begin{equation}
    \sum_{l=0}^N c_l \bra{\frac{N}{2}, -\frac{N}{2}} R^{-1}_s \ket{\frac{N}{2}, l - \frac{N}{2}} = \sum_{l=0}^N c_l (-1)^l \sqrt{C^N_l} \left( \cos{\left( \frac{\beta_s}{2} \right)} \right)^{N-l} \left( \sin{\left( \frac{\beta_s}{2} \right)} \right)^l e^{i \left( l - \frac{N}{2} \right) \alpha_s} = 0.
\end{equation}

By simplifying, we obtain the polynomial form:

\begin{equation}
    P(\epsilon) = \sum_{l=0}^N c_l (-1)^l \sqrt{C^N_l} \epsilon^l = 0,
\end{equation}
where \( \epsilon_s = \tan{\left( \frac{\beta_s}{2} \right)} e^{i \alpha_s} \). Given the coefficients \( c_l \), the \( N \)-roots \( \epsilon_s \) of this polynomial determine the orientations \( \left( \alpha_s, \beta_s \right) \) of the spinors constituting the \( N \)-qubit symmetric state. This leads to an intrinsic geometric picture of the system in terms of \( N \)-points on the unit sphere \( S^2 \) \cite{Majorana1932, Hannay1998, Ribeiro2007}.

In other words, using an inverse stereographic projection, we can find the orientation of the spin on the Bloch sphere:

\begin{equation}
    \ket{\epsilon_s} = \cos{\left(\frac{\beta_s}{2}\right)} e^{-i \frac{\alpha_s}{2}} \ket{0} + \sin{\left(\frac{\beta_s}{2}\right)} e^{i \frac{\alpha_s}{2}} \ket{1}.
\end{equation}

Thus, the geometric interpretation of a symmetric \(N\)-qubit state is that it corresponds to \(N\) points, or Majorana stars, on the surface of the Bloch sphere. These points encapsulate the essential features of the state, providing a compact and visually interpretable representation of the system’s structure. This connection between the symmetric state and the Majorana representation plays a significant role in quantum information theory, allowing us to view multi-qubit states geometrically \cite{Bastin2009, Markham2011}.

The orientations of the spinors \( \left( \alpha_s, \beta_s \right) \) corresponding to the roots of the polynomial \( P(\epsilon) \) provide a complete description of the symmetric state, making the Majorana representation a powerful tool for analyzing and visualizing the properties of quantum states.

\section{Galois Groups and Quantum State Classification}

In this section, we explore the connection between Galois groups of polynomials and the classification of symmetric three-qubit quantum states using the Majorana representation. By examining polynomials of degree three with complex coefficients, we demonstrate how the structure of their Galois groups correlates with the entanglement properties and symmetries of the corresponding quantum states \cite{Chryssomalakos2017}.

An arbitrary symmetric three-qubit state can be associated with a cubic polynomial whose roots determine the orientations of the qubits on the Bloch sphere. The polynomial is given by:

\begin{equation}
P(\epsilon) = c_0 - c_1 \sqrt{3}\, \epsilon + c_2 \sqrt{3}\, \epsilon^2 - c_3\, \epsilon^3 = 0,
\end{equation}

where \( c_l \) are the coefficients of the state's expansion in the Dicke basis \( \{ \ket{\tfrac{3}{2}, m} \} \) with \( m = -\tfrac{3}{2}, -\tfrac{1}{2}, \tfrac{1}{2}, \tfrac{3}{2} \). The roots \( \{ \epsilon_i \} \) of \( P(\epsilon) \) correspond to the stereographic projections of the qubit orientations in the Majorana representation.

We focus on the \textbf{symmetries among the roots of polynomials}, much like the method pioneered by Galois. By examining these symmetries, we delve into the intrinsic relationships and structural patterns that exist among the roots, which are pivotal for understanding phenomena such as quantum entanglement \cite{Stewart2004}. Galois's work reveals that these symmetries can be captured by specific groups—now known as \textbf{Galois groups}—which encapsulate the permutations of roots that leave the fundamental algebraic equations unchanged. By analyzing these groups, we can classify and comprehend the properties of quantum states, as the symmetries among the roots directly correlate with the entanglement and other quantum characteristics \cite{Bengtsson2006}.

To investigate the symmetries underlying the entanglement properties of the three-qubit GHZ state, we consider the polynomial equation:
\begin{equation}
P(\epsilon) = \epsilon^3 - 1 = 0.
\end{equation}
This equation has as its solutions the three cube roots of unity:
\begin{equation}
\epsilon_1 = 1, \quad \epsilon_2 = e^{i \frac{2\pi}{3}}, \quad \epsilon_3 = e^{i \frac{4\pi}{3}}.
\end{equation}
The splitting field of \( P(\epsilon) \) is \( \mathbb{Q}(\epsilon) \), with \( \epsilon \) being a primitive cube root of unity. The Galois group \( \mathrm{Gal}(P) \) is isomorphic to the cyclic group \( C_3 \), generated by the automorphism: $\sigma: \epsilon \mapsto \epsilon^2.$ This group encompasses the cyclic permutations (rotations) of the roots, embodying the full rotational symmetry \( C_3 \) \cite{Artin2015}.

By selecting the coefficients \( c_0 = 1 \), \( c_1 = 0 \), \( c_2 = 0 \), and \( c_3 = 1 \), the polynomial can be rewritten as:
\begin{equation}
P(\epsilon) = 1 - \epsilon^3 = 0.
\end{equation}
This corresponds to the symmetric three-qubit GHZ state \cite{Greenberger1990}:
\begin{equation}
\ket{\mathrm{GHZ}} = \frac{1}{\sqrt{2}} \left( \ket{000} + \ket{111} \right) .
\end{equation}

The cyclic symmetry group \( C_3 \) corresponds to the permutation symmetry of the GHZ state under cyclic exchanges of qubits. Specifically, \( C_3 \) consists of three elements: the identity and two cyclic permutations. The rotations correspond to cyclic permutations of the roots, reflecting the high symmetry of the GHZ state \cite{Dur2000}.

Geometrically, the roots \( \epsilon_1, \epsilon_2, \epsilon_3 \) of a maximally entangled GHZ state are located at the vertices of an equilateral triangle inscribed in the unit circle of the complex plane. When mapped onto the Bloch sphere via the Majorana representation, each root corresponds to a Majorana star on the sphere's equator. The equidistant arrangement of these stars forms an equilateral triangle on the Bloch sphere, reflecting the \( C_3 \) symmetry \cite{Bastin2009}.

This symmetric distribution of the Majorana stars signifies maximal tripartite entanglement of the GHZ state. The cyclic symmetry ensures that the entanglement properties are invariant under any permutation of the qubits, highlighting the state's robust symmetric properties and uniform entanglement among all qubits in the system. Breaking this symmetry—thus reducing the symmetry group from \( C_3 \) to \( C_2 \)—leads to a loss of entanglement in the system. In general, losing any symmetry with respect to the center of the Bloch sphere results in diminished entanglement, as the uniform distribution of the Majorana stars is disrupted \cite{Markham2011}.

Furthermore, if one of the vertices of the equilateral triangle is moved, some of the symmetries are lost. As this vertex continues to move and eventually coincides with another vertex, the symmetry group reduces even further. When two roots become identical, the system's symmetry breaks from \( C_3 \) to \( C_2 \), since we now have three roots with two of them being the same. This reduction in symmetry corresponds to a further decrease in entanglement. In this scenario, the resulting quantum state corresponds to the W state in a three-qubit system \cite{Dur2000}:
\begin{equation}
\ket{W} = \frac{1}{\sqrt{3}} \left( \ket{001} + \ket{010} + \ket{100} \right) .
\end{equation}
The W state exhibits a different type of entanglement compared to the GHZ state. While the GHZ state loses all entanglement when any qubit is traced out, the W state retains bipartite entanglement upon tracing out a qubit. The robustness of the W state arises from the presence of double roots, resulting in only two distinct Majorana stars on the Bloch sphere \cite{Benz1}. This configuration correlates similarly to a two-qubit system but offers enhanced robustness due to the duplication of one root \cite{Wei2003}. This suggests that increasing the number of qubits and introducing degenerate roots can further enhance the robustness of the entanglement. By creating states with multiple degenerate roots, one can potentially maintain higher levels of entanglement even under perturbations, highlighting the critical role that geometric and algebraic symmetries play in sustaining quantum correlations within larger multipartite systems \cite{Markham2011}.

Finally, if we continue to move the third vertex until it coincides with the other two, all three roots merge into a single triple root:
\begin{equation}
\epsilon_1 = \epsilon_2 = \epsilon_3 = 1.
\end{equation}
In this case, the polynomial becomes:
\begin{equation}
P(\epsilon) = (1 - \epsilon)^3 = 0,
\end{equation}
which represents a fully degenerate root. The corresponding quantum state is the fully separable state:
\begin{equation}
\ket{\Psi_{\mathrm{Sep}}} = \ket{\epsilon,\epsilon,\epsilon}.
\end{equation}
This separable state exhibits no entanglement, as each qubit is in a definite state independently of the others. Geometrically, all Majorana stars collapse to a single point on the Bloch sphere's equator, eliminating any geometric distribution that could facilitate entanglement \cite{Bastin2009}.

In this fully degenerate scenario, all symmetries except the trivial identity are broken. The symmetry group reduces to the identity group \( \{e\} \), as there are no non-trivial permutations that leave the single root unchanged. Mathematically, this is expressed as:
\begin{equation}
\mathrm{Gal}(P) \cong \{e\},
\end{equation}
indicating that the only automorphism is the identity map \cite{Stewart2004}. With the complete loss of symmetry, the state lacks any entanglement, underscoring the essential role that symmetries play in maintaining quantum correlations. This transition from maximal symmetry and entanglement in the GHZ state, through the W state with partial symmetry, to the fully separable state with no symmetry, illustrates the delicate interplay between geometric configurations, symmetry groups, and the entanglement properties of multipartite quantum systems \cite{Chryssomalakos2017}. Other classes of states exhibit partial separability but cannot be expressed as pure symmetric states; their representation and quantification have been thoroughly investigated in previous works \cite{Benz2}.

\section{Depressed Cubic Polynomials and Their Implications for Quantum States}

To further explore the relationship between polynomial symmetries and quantum entanglement, we analyze the general cubic polynomial associated with a three-qubit symmetric state. Consider the polynomial:
\begin{equation}
P(\epsilon) = c_0 - c_1 \sqrt{3}\, \epsilon + c_2 \sqrt{3}\, \epsilon^2 - c_3\, \epsilon^3 = 0.
\end{equation}

This polynomial corresponds to the symmetric three-qubit state:
\begin{equation}
\ket{\Psi}=c_0 \ket{000}+\frac{c_1}{\sqrt{3}} ( \ket{001}+\ket{010}+\ket{100} )+\frac{c_2}{\sqrt{3}} (\ket{011}+\ket{101}+\ket{110})+c_3\ket{111}
\end{equation}

To simplify the analysis, we transform \( P(\epsilon) \) into its depressed cubic form, which eliminates the quadratic term. This is achieved by substituting \( \epsilon = y - \frac{c_2}{c_3\sqrt{3}} \), leading to:
\begin{equation}
P_{\text{depressed}}(y) = y^3 + p y + q = 0,
\end{equation}
where the coefficients \( p \) and \( q \) are expressed in terms of the original coefficients \( c_0, c_1, c_2, \) and \( c_3 \) \cite{Abramowitz1964}:
\begin{align}
p &= \frac{c_3 c_0 - c_2^2}{c_3^2}, \\
q &= \frac{2 c_2^3 - 3 c_3 c_0 c_2 + 9c_3^2 c_1}{3\sqrt{3}c_3^3}.
\end{align}

The conditions for the nature of the roots are determined by the values of \( p \) and \( q \). Specifically:
\begin{itemize}
    \item \textbf{Separable State (Triple Root)}: 
    
    When both \( p = 0 \) and \( q = 0 \), the depressed cubic simplifies to \( y^3 = 0 \), yielding a triple root \( y = 0 \). Substituting back, this corresponds to all three roots \( \epsilon_1 = \epsilon_2 = \epsilon_3 = \epsilon_0 \), leading to the fully separable state:
    \begin{equation}
    \ket{\Psi_{\mathrm{Sep}}} = \ket{000}.
    \end{equation}
    In this case, the polynomial becomes:
    \begin{equation}
    P(\epsilon) = (1 - \epsilon)^3 = 0,
    \end{equation}
    indicating complete degeneracy with all Majorana stars collapsing to a single point on the Bloch sphere. The symmetry group reduces to the trivial identity group \( \{e\} \), and the state exhibits no entanglement \cite{Bastin2009}.

    \item \textbf{GHZ State (Triple Root)}: 
    
    When \( p = 0 \) but \( q \neq 0 \), the depressed cubic takes the form \( y^3 + q = 0 \), resulting in three distinct roots with a $C_3$ symmetry. This scenario corresponds to three distinct roots \( \epsilon_1, \epsilon_2, \epsilon_3 \) arranged symmetrically around the origin in the complex plane, leading to a maximum entangled GHZ state when $|q|= 1$ \cite{Greenberger1990}:
    \begin{equation}
    \ket{\mathrm{GHZ}} = \frac{1}{\sqrt{2}} \left( \ket{000} + \ket{111} \right).
    \end{equation}
    The polynomial in this case is:
    \begin{equation}
    P(\epsilon) = (1 - \epsilon)(1 - \omega \epsilon)(1 - \omega^2 \epsilon) = 0,
    \end{equation}
    where \( \omega = e^{2\pi i /3} \) is a primitive third root of unity. The roots exhibit rotational symmetry, reflecting the high entanglement and symmetry of the GHZ state. The symmetry group is \( C_3 \), corresponding to cyclic permutations of the roots \cite{Stewart2004}.

    \item \textbf{W State (Single Root and Double Root)}: 
    
    The W state arises when the depressed cubic has one distinct root and a double root, corresponding to two coinciding Majorana stars and one distinct Majorana star on the Bloch sphere. This scenario does not correspond to \( p = 0 \) but instead requires specific relationships between \( p \) and \( q \) to achieve the necessary root structure. The corresponding state is \cite{Dur2000}:
    \begin{equation}
    \ket{W} = \frac{1}{\sqrt{3}} \left( \ket{001} + \ket{010} + \ket{100} \right) .
    \end{equation}
    The polynomial in this case is:
    \begin{equation}
    P(\epsilon) = (\epsilon - \epsilon_1)^2 (\epsilon - \epsilon_3) = 0,
    \end{equation}
    where \( \epsilon_1 \) and \( \epsilon_3 \) are distinct. The presence of a double root results in two coinciding Majorana stars on the Bloch sphere, correlating similarly to a two-qubit system but with enhanced robustness due to the duplication of one root \cite{Wei2003}. The symmetry group is reduced to \( C_2 \), reflecting the partial symmetry of the state.
\end{itemize}

Transforming our cubic polynomial (20) into a depressed one (22) is similar to applying a unitary transformation, the symmetric state can be reduced to eliminate certain excitation terms, effectively compressing the information contained within the state. For the three-qubit case, the state simplifies to:
\begin{equation}
\ket{\Psi} = c_0' \ket{000} + \frac{c_1'}{\sqrt{3}} \left( \ket{001} + \ket{010} + \ket{100} \right) + c_3' \ket{111},
\end{equation}

The transformation of coefficients $c_0'$ and $c_3'$ demonstrates that the information contained in the two-excitation subspace ($\ket{011} + \ket{101} + \ket{110}$) can be effectively represented within the single-excitation and three-excitation subspaces without loss of generality \cite{Mathonet2010}. This unitary transformation enables the representation of any 3-qubit symmetric state as a superposition of $\ket{GHZ}$, $\ket{W}$, and $\ket{\Psi_{Sep}}$ states:
\begin{equation}
\ket{\Psi} = \frac{1}{\sqrt{N}}(c_0' \ket{GHZ} + c_1' \ket{W} + (c_3'-c_0') \ket{\Psi_{Sep}}),
\end{equation}
where $N$ is the normalization factor. Given the orthonormality of the $\ket{GHZ}$ and $\ket{W}$ states, we can construct a measure of entanglement for 3-qubit symmetric states based on the sum of the squared magnitudes of coefficients $c_0'$ and $c_1'$:
\begin{equation}
D_{ent} = \frac{2}{N}(|c_0'|^2 + |c_1'|^2).
\end{equation}
The maximum entanglement occurs when $c_0' = 1$, resulting in a pure GHZ state. In this case, $D_{ent}(GHZ) = 1$. Conversely, when $c_1' = 1$, we obtain a pure W state, yielding $D_{ent}(W) = \frac{2}{3}$, which is lower than the entanglement measure for the GHZ state, consistent with the known properties of these multipartite entangled states \cite{Wei2003}.

\section{Depressed Quartic Polynomials and Their Impact on Four-Qubit Symmetric States}

Extending our analysis to four-qubit symmetric states, we consider the general quartic polynomial associated with such states:
\begin{equation}
P(\epsilon) = c_4 \epsilon^4 + c_3 \epsilon^3 + c_2 \epsilon^2 + c_1 \epsilon + c_0 = 0.
\end{equation}
This polynomial corresponds to the symmetric four-qubit state:
\begin{equation}
\ket{\Psi} = c_0 \ket{0000} + c_1 \ket{\mathrm{D}^{(1)}} + c_2 \ket{\mathrm{D}^{(2)}} + c_3 \ket{\mathrm{D}^{(3)}} + c_4 \ket{1111},
\end{equation}
where \( \ket{\mathrm{D}^{(k)}} \) denotes the symmetric Dicke state with \( k \) excitations \cite{Dicke1954}:
\begin{equation}
\ket{\mathrm{D}^{(k)}} = \frac{1}{\sqrt{\binom{4}{k}}} \sum_{\text{permutations}} \ket{\underbrace{1 \cdots 1}_{k} \underbrace{0 \cdots 0}_{4-k}} .
\end{equation}

To simplify the quartic polynomial, we perform a change of variable to eliminate the \emph{quadratic} term (the $\epsilon^2$ term). We set $\epsilon = y - \frac{c_2}{4 c_4}$. Substituting this into $P(\epsilon)$ yields the depressed quartic polynomial $y^4 + p y^2 + q y + r = 0$, where the coefficients $p$, $q$, and $r$ are given by:

\begin{align}
p &= \frac{3 c_2^2 - 8 c_1 c_4}{8 c_4^2}, \\
q &= \frac{c_2^3 - 4 c_1 c_2 c_4 + 8 c_0 c_4^2}{8 c_4^3}, \\
r &= \frac{256 c_4^3 c_0 - 64 c_4^2 c_1 c_2 + 16 c_4 c_2^3 - 3 c_2^4}{256 c_4^4}.
\end{align}

Importantly, eliminating the quadratic term in the depressed quartic polynomial corresponds to removing the three-excitation term \( c_3 \ket{\mathrm{D}^{(3)}} \) in the symmetric four-qubit state. There exists a unitary transformation \( U \) that achieves this compression:
\begin{equation}
U \ket{\Psi} = c_0' \ket{0000} + c_1' \ket{\mathrm{D}^{(1)}} + c_2' \ket{\mathrm{D}^{(2)}} + c_4' \ket{1111},
\end{equation}
where the transformed coefficients \( c_0' \), \( c_1' \), \( c_2' \), and \( c_4' \) encapsulate the original state's information without the three-excitation component. Analyzing the conditions on \( p \), \( q \), and \( r \) provides insights into the entanglement properties of the corresponding quantum state:

\textbf{Separable State (All roots equal):}

$p = 0$, \quad $q = 0$, \quad $r = 0$. The depressed quartic simplifies to $y^4 = 0$, indicating a quadruple root and corresponding to the fully separable state $\ket{\Psi_{\mathrm{Sep}}} = \ket{0000}$.

\textbf{GHZ State (Maximal entanglement):}

We set \( p = 0 \), \( q = 0 \), and \( r \neq 0 \). The depressed quartic becomes \( y^4 + r = 0 \), and for specific values of \( r \) (e.g., \( |r| = 1 \)), the roots are symmetrically distributed, reflecting maximal entanglement in the GHZ state \cite{Greenberger1990}:
\[
\ket{\mathrm{GHZ}_4} = \frac{1}{\sqrt{2}} \left( \ket{0000} + \ket{1111} \right).
\]

\textbf{Two Double Roots (Partial entanglement):}

$q = 0, \quad r = \frac{p^2}{4}$. The depressed quartic factors as $\left( y^2 + \frac{p}{2} \right)^2 = 0$, indicating two double roots. This corresponds to a state resembling a W state with two excitations: 
$\ket{W_{2 \times 2}} = \frac{1}{\sqrt{6}} \left( \ket{y_1,y_1,y_2,y_2} + \text{permutations} \right)$, 
where the permutations include all distinct states with two excitations. The symmetry group is $C_2$, similar to a standard W state with two points on the Bloch sphere. The presence of double roots enhances the state's robustness, as the duplication of roots contributes to resilience against decoherence, analogous to the robustness observed in three-qubit W states \cite{Dur2000}.

\textbf{W State Analogue (Triple root and one distinct root):}

  Achieving a triple root and one distinct root requires specific relationships among \( p \), \( q \), and \( r \), leading to a depressed quartic of the form \( (y - y_1)^3 (y - y_2) = 0 \). The corresponding quantum state resembles the four-qubit W state:
  \begin{equation}
  \ket{W_4} = \frac{1}{2} \left( \ket{y_1,y_1,y_1,y_2} + \text{permutations} \right),
  \end{equation}
  which can be a maximally entangled W state when $y_1=0$ and $y_2=1$.
  The symmetry group remains \( C_2 \), and the presence of a triple root enhances the state's robustness, similar to the three-qubit W state with double roots. The duplication of roots in this case also contributes to the state's resilience, indicating that the entanglement is preserved under particle loss \cite{Wei2003}.

The elimination of the three-excitation term via the unitary transformation and the removal of the quadratic term in the depressed quartic are intrinsically linked. This compression simplifies the state's structure while preserving essential entanglement properties, allowing for a more straightforward analysis of the system.

By manipulating the coefficients \( c_0 \) through \( c_4 \) to satisfy the conditions on \( p \), \( q \), and \( r \), we can design quantum states with desired entanglement characteristics. This demonstrates the profound connection between the algebraic structure of polynomials and quantum entanglement in multipartite systems \cite{Chryssomalakos2017}.

Following the same approach as in the 3-qubit case, we can express the state as a superposition of GHZ, $W_4$, $W_{2\times2}$, and separable states, which can be written as

\begin{equation} \ket{\Psi} = \frac{1}{\sqrt{N}} \left( c_0' \ket{GHZ} + c_1' \ket{W_4} + c_2' \ket{W_{2\times2}} + (c_4' - c_0') \ket{\Psi_{\text{Sep}}} \right), \end{equation}

where $N$ is a normalization factor, and $c_0'$, $c_1'$, $c_2'$, and $c_4'$ are the corresponding complex coefficients. The entanglement measure, $D_{\text{ent}}$, is defined as the sum of the squared norms of the coefficients associated with the non-separable states (GHZ, $W_4$, and $W_{2\times2}$), leading to

\begin{equation} D_{\text{ent}} = \frac{2}{N} \left( |c_0'|^2 + |c_1'|^2 + |c_2'|^2 \right). \end{equation}

Notably, the contributions to entanglement follow a clear hierarchy, with the GHZ component contributing the most, followed by the $W_4$ and $W_{2\times2}$ states:

\begin{equation} D_{\text{ent}}(\text{GHZ}) > D_{\text{ent}}(W_4) > D_{\text{ent}}(W_{2\times2}). \end{equation}

This decomposition highlights the relative significance of each entangled component, with the GHZ state dominating the overall entanglement, providing a valuable framework for analyzing complex multi-qubit systems. The clear hierarchy of contributions offers insights into how different types of multipartite entanglement interact in larger systems, which could have implications for quantum information processing tasks \cite{Guhne2009}.

\section{Conclusion}
In this study, we have introduced a new method to quantify entanglement in multi-qubit symmetric states by leveraging the algebraic structures of Galois groups. This approach not only provides a fresh interpretation of quantum entanglement classification but also offers a practical and scalable measure that can be easily generalized to systems with an arbitrary number of qubits. By framing the classification of GHZ, $W$, and separable states within Galois symmetries, we reveal a hierarchy in their entanglement properties that can be systematically analyzed across larger quantum systems.

The elegance of this method lies in its simplicity and generalizability, making it applicable to a wide range of multi-qubit configurations without the computational overhead typically associated with entanglement measures. This opens up new possibilities for both theoretical studies and practical implementations in quantum information science, where efficient quantification of entanglement is crucial for developing quantum technologies \cite{Amico2008}.

As quantum systems continue to scale, the ability to quantify entanglement in a structured and systematic way will become increasingly essential. Our findings not only bridge an important gap between algebraic theory and quantum mechanics but also point to future research directions where the Galois framework can be extended to more complex quantum systems. The insights from this work have the potential to influence how we approach quantum state classification and optimization in quantum computing, with broad implications for quantum communication and cryptography \cite{Guhne2009}.

\end{document}